# The Origins Space Telescope:
# Towards An Understanding of Temperate Planetary Atmospheres

Jonathan J. Fortney, University of California, Santa Cruz, jfortney@ucsc.edu, 831-459-1312

Tiffany Kataria, Jet Propulsion Laboratory - California Institute of Technology
Kevin Stevenson, Space Telescope Science Institute
Robert Zellem, Jet Propulsion Laboratory - California Institute of Technology
Eric Nielsen, Kavli Institute for Particle Astrophysics and Cosmology, Stanford University
Pablo Cuartas-Restrepo, FACom-SEAP, FCEN, Universidad de Antioquia, Medellín, Colombia
Eric Gaidos, University of Hawaii at Manoa
Edwin Bergin, University of Michigan, Ann Arbor
Margaret Meixner, Space Telescope Science Institute
Stephen Kane, University of California, Riverside
David Leisawitz, Sciences and Exploration Directorate, NASA Goddard Space Flight Center
Jonathan Fraine, Space Telescope Science Institute
Lisa Kaltenegger, Carl Sagan Institute, Cornell University
Angelle Tanner, Mississippi State University
Mercedes Lopez-Morales, Harvard-Smithsonian Center for Astrophysics
Tom Greene, NASA Ames Research Center
William Danchi, NASA Goddard Space Flight Center
Keivan Stassun, Vanderbilt University
Ravi Kopparapu, NASA Goddard Space Flight Center/UMCP
Eric Wolf, University of Colorado, Boulder
Tiffany Meshkat, IPAC - California Institute of Technology
Natalie Hinkel, Vanderbilt University
Klaus Pontoppidan, Space Telescope Science Institute
Chuanfei Dong, Department of Astrophysical Sciences, Princeton University
Giovanni Bruno, Space Telescope Science Institute
Dawn Gelino, NASA Exoplanet Science Institute
Vladimir Airapetian, NASA Goddard Space Flight Center/American University
Eric Agol, University of Washington, Virtual Planetary Laboratory, Guggenheim Fellow
Drake Deming, University of Maryland
Jacob Haqq-Misra, Blue Marble Space Institute of Science
Niki Parenteau, NASA Ames Research Center
Carey Lisse, Johns Hopkins University Applied Physics Laboratory
Gregory Tucker, Brown University
Prabal Saxena, NASA Goddard Space Flight Center
Robin Wordsworth, Harvard University
Geoffrey Blake, Division of Geological & Planetary Sciences, California Institute of Technology
Shannon Curry, University of California, Berkeley
Zachory Berta-Thompson, University of Colorado Boulder
Malcolm Fridlund, University of Leiden
Kate Su, University of Arizona
Peter Gao, University of California, Berkeley
Vardan Adibekyan, Instituto de Astrofísica e Ciências do Espaço, Universidade do Porto
Nicholas Heavens, Department of Atmospheric and Planetary Sciences, Hampton University
Dante Minniti , Dept. of Physical Sciences, Universidad Andres Bello, Chile
Sarah Rugheimer, Centre for Exoplanet Science, University of St. Andrews
Benjamin Rackham, Department of Astronomy / Steward Observatory, University of Arizona
Kathleen Mandt, Johns Hopkins University Applied Physics Laboratory
Miguel de Val-Borro, NASA Goddard Space Flight Center
Tyler Robinson, Northern Arizona University

The past 30 years of exoplanet science have shown us that planets are common around nearby stars. Now and in the coming years, we would like to characterize these worlds. What are the atmospheres of temperate rocky planets like? Do these planets have life? Can we understand the continuum of all kinds of atmospheres, from hot to warm to cool? The *Origins Space Telescope (OST)* will expand upon the legacy of infrared exoplanet science with *Spitzer*, *Hubble*, and the upcoming *James Webb Space Telescope* (*JWST*) by conducting transmission and emission (dayside and phase-resolved) spectroscopy of transiting exoplanets, with a particular focus on Earth-size exoplanets transiting in the habitable zones of mid-to-late M-dwarf stars. Furthermore, *OST* coronagraphic investigations will yield detections of true Jupiter and Saturn analogs. Ultimately, *OST* will probe exoplanet atmospheres with a large range of physical characteristics and over a broad wavelength range that will inevitably lead to many unanticipated scientific discoveries and, with any luck, a highly anticipated one -- the signature of life.

*OST* is the mission concept for the Far-Infrared Survey Study, one of four mission concepts currently being studied by NASA in preparation for the Astrophysics 2020 Decadal Survey. With active cooling to 4 K, *OST* will be extremely sensitive in mid- to far-IR wavelengths from 5-600 μm. It will use imaging and spectroscopy to probe the furthest reaches of our known universe, trace the path of water through star and planet formation, and place thermochemical constraints on the atmospheres of exoplanets ranging in size from Jupiter to Earth. Mission Concept 1, featuring a 9.1m primary mirror, will be the focus of the OST 2018 Interim Report. Concept 2, featuring a 5.9m primary (the same collecting area as JWST) will be the focus of study throughout 2018. This contribution to the *Exoplanet Science Strategy* panel discusses the significant advancements that the *OST Mid-Infrared Imager, Spectrometer, and Coronagraph (MISC)* instrument can make in studying cool planetary atmospheres.

**Advantages of the Transit Technique**

Built on its experience with *Spitzer*, *Hubble*, and *JWST*, the exoplanet community has developed a successful tradition of using the transit technique to probe the physics and chemistry of exoplanet atmospheres. Transit measurements have paid dividends for nearly 20 years (e.g., Charbonneau et al. 2002, Sing et al. 2016), and *OST* can build on these strengths. While these dividends are surely due to the large number of discoveries, this science is also aided significantly by the physical properties constrained from transits, namely the planetary radius, and when combined with a mass measurement, its bulk density. These two fundamental parameters are the essential first step toward identifying rocky, potentially habitable worlds and neither value can be unambiguously constrained using other techniques.Using this knowledge of bulk density, we can point *OST* at those planets that are most likely to be rocky, Earth-like and have atmospheres that could contain signs of life.

Distinguishing potentially habitable planets from those more likely to have inhospitable surfaces requires information about their atmospheres. During primary transit, a small fraction of the light from the host star passes through the planet's atmosphere, allowing *OST* to measure its transmission spectrum (manifest as a wavelength-dependent radius). This method has been used to successfully detect absorption by atoms and molecules in the atmospheres of hot, Jupiter-size planets (e.g., Charbonneau et al., 2002; Sing et al., 2016) Furthermore, transmission spectra can yield the pressure-altitude of any clouds, important information on condensable species. What previous ground- and space-based telescopes did for hot Jupiters, *OST* will achieve for temperate Earth-size exoplanets.

**Advantages of the Thermal Infrared**

During conjunction or secondary eclipse the planet passes behind its host star and by difference, *OST* will measure the its thermal emission. The emission from a planet in the habitable zone (HZ), as well as the planet-to-star flux ratio, peaks in the mid-infrared. Moreover, the limiting effect of noise from star spots and stellar granulation on the transmission spectrum is minimized in the mid-infrared (Trampedach et al. 2013, Rackham et al. 2018). This secondary eclipse spectrum probes molecular species as well as the temperature vs. altitude profile of the planet's dayside atmosphere (e.g., Kreidberg et al., 2014; Line et al., 2016). For instance, Earth's emission spectrum shows that our planet has a temperature inversion; the strongest part of the $CO_2$ band at 15 μm, formed at high altitudes, is seen in emission, rather than absorption.

Outside of secondary eclipse, broadband spectrophotometric observations will be sufficiently sensitive to measure phase-resolved thermal emission of terrestrial planets (see the *OST* White Paper by Kataria et al. 2018). The magnitude of the day-night temperature contrast will constrain the redistribution of heat and hence the thickness and circulation of the atmosphere. Because these phase observations will be conducted spectroscopically, both longitude and altitude will be probed, revealing for the first time the global thermal structure of temperate planet atmospheres. Furthermore, determination of day- and night-side thermal fluxes from observations at different phases, combined with the incident irradiance, yields the planetary Bond albedo (e.g., Stevenson et al. 2014).

High-cadence observations during eclipse ingress and egress, a technique known as eclipse mapping, produces a "raster scan" of the planet's day side emission. This has been applied to infer information about atmospheric circulation on hot Jupiters (Majeau et al. 2012, de Wit et al. 2012) and will be practical for smaller and cooler planets with *OST*. Planet-planet occultations provide a further probe of planets at different phases, and may be amenable to study with *OST* for short-period planets such as in the TRAPPIST-1 system (Luger et al. 2018).

**Detecting Biosignatures in Temperate Earth-Sized Planets Transiting Mid/Late M Stars**

By the mid 2030s, TESS (Sullivan et al. 2015), MEarth (Berta-Thompson et al. 2013), TRAPPIST (Gillon et al. 2016), SPECULOOS (Gillon 2017), and future surveys will have systematically searched all nearby M-dwarf stars and identified dozens of transiting habitable-zone exoplanets. Many of those orbiting mid-to-late M dwarfs will have well-determined masses using next-generation Doppler radial velocity instruments and/or transit timing variations. Knowing the mass and radius provides an estimate of a planet's bulk density and volatile inventory. An example of this is shown in Figure 10 of Grimm et al. (2018), where six of the seven TRAPPIST-1 planets have bulk densities that are consistent with volatile-rich envelopes in the form of thick atmospheres, oceans, and/or ice.

Atmospheres of temperate terrestrial (and giant) planets are predicted to contain numerous prominent molecular absorption features in the mid-infrared. Examples include carbon dioxide ($CO_2$, 15 μm), water vapor ($H_2O$, 6.3 and 18 μm), ozone ($O_3$, 9.6 μm), and methane ($CH_4$, 7.7 μm). Less prominent molecules include nitrous oxide ($N_2O$, 16.9 μm), ammonia ($NH_3$, 11 μm), chloromethane ($CH_3Cl$), and a host of methane-derived photochemical products (e.g. $C_2H_2$, $C_2H_4$, $C_2H_6$, etc.). Anthropogenic gases like chloro-fluorocarbons ($CCl_2F_2$ and $CCl_3F$) also have strong absorption features in the mid-infrared.

The mid-infrared is rich in molecular features for biologically interesting gases, but for an Earth-like biosphere, two important detectable constituents that could indicate the presence of life are the combination of $O_3$ and $CH_4$ (e.g. Kaltenegger, 2017). A significant strength of *OST* is its broad simultaneous wavelength coverage (5-25 μm) and, thus, its ability to efficiently detect multiple molecules and put them into context (see Figure 1). For instance, in Venus the absence of water vapor and the ubiquity of carbon dioxide is seen as a signpost of a runaway greenhouse effect that is inhospitable for life. Furthermore, most molecules can be probed in both transmission and emission, thus giving a planet-wide view of their abundances.

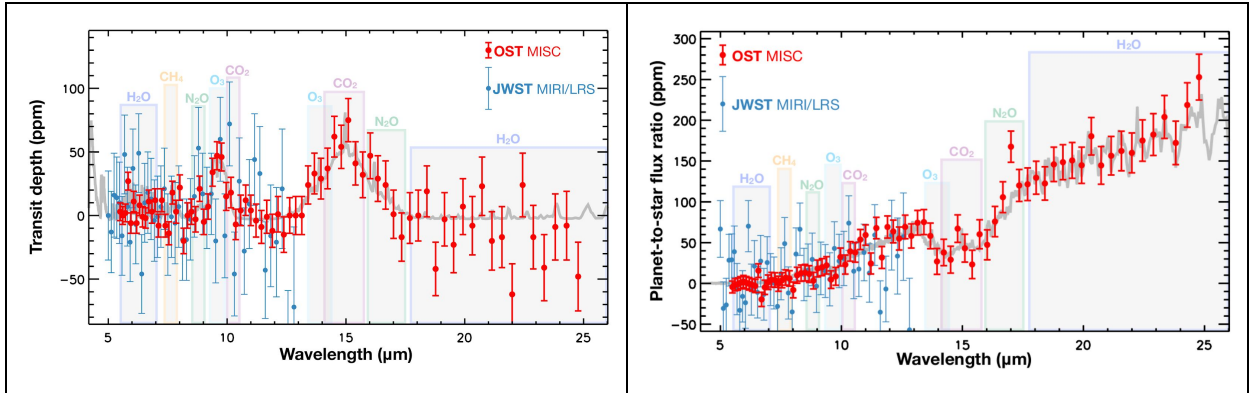

Figure 1. *OST* will characterize habitable planets in search of signs of life. Model transmission (left) and emission (right) spectra of TRAPPIST-1e (0.92 $R_E$, 250 K) with simulated JWST (blue) and *OST* (red) data. The colored panels indicate bandwidths of detectable molecular features. Uncertainties are derived for a $K_{mag}$ = 8 star using 30 transits/eclipses. The JWST+MIRI/LRS simulations assume an optimistic noise floor of 30 ppm. The *OST* simulations assume a 9-meter aperture and no noise floor. A noise floor of 5 ppm looks similar.

**How common are terrestrial planets orbiting M dwarfs?**

The expected sample size of terrestrial planets that transit in the HZ of mid-to-late M dwarfs is not yet well-known. Parent stars that will yield the best signal-to-noise observations will be within ~15 pc. Estimates of planet frequency from the Kepler Mission for M stars are hampered by the fact that nearly all Kepler M stars are early M (e.g. Dressing & Charbonneau, 2015). Currently, four sub-15 pc transiting HZ exoplanets are known, including TRAPPIST-1 d, e, and f (Gillon et al. 2017) and LHS 1140 b (Dittman et al. 2017). Several additional non-transiting planets are also known, including Proxima-b and Ross 128b. A projection of the Sullivan et al. (2015) sample for TESS -- again based on the Kepler statistics, and thus on early M dwarfs -- suggests ~8 temperate rocky planets (< 300 K, < 1.5 $R_E$) around M stars less than 0.25 $M_{Sun}$. Furthermore, the ground-based project SPECULOOS (Gillon 2017) is anticipated to detect an additional 25 HZ planets, a few of which will be within 15 pc, for a total of at least a dozen quality targets. However, there are at least two reasons that this number could be a factor of ~2 too low. Quantitatively, Ballard (2018) utilize a dual-population planet occurrence model, which includes compact multiple planetary systems, to infer a revised TESS planet yield for early M dwarfs that is 50% higher than that of Sullivan et al. (2015). And qualitatively, the detection of the TRAPPIST-1 system suggests that unless this was a "lucky" detection in the first 50 systems of that survey, the frequency of HZ planets around such stars could be higher. Therefore, a number closer to ~two dozen seems quite plausible.

**M Dwarfs as Habitable Planet Hosts**

It is well established that M stars make up at least 75% of all stars in our solar neighborhood (and likely the galaxy). As such, M-dwarf planetary systems are likely the main

"mode" of planet formation. They also dominate in terms of the number of Earth-size planets in the HZ of their parent stars (Dressing & Charbonneau, 2015, Gaidos et al. 2016). There has been significant discussion in the literature concerning the viability of planets around M dwarfs as suitable abodes for life due to potential tidal locking and initially high luminosity in the formation phase. Tidal locking has been shown to not be a serious concern for the temperature difference on the day and night sides of planets (e.g. Joshi et al 1997) and slow rotation could even increase the HZ limit due to cloud coverage on the dayside (Kopparapu et al. 2017). High bolometric luminosity for these stars in their pre-main sequence phase (e.g., Ramirez & Kaltenegger 2014, Luger & Barnes 2015), as well as the high ratio of $L_{XUV}/L_{BOL}$ for M stars compared to Sun-like stars, could initially leave planets volatile-poor and subject to extreme atmospheric evaporation (Airapetian et al. 2017, Dong et al. 2017). However processes similar to the late heavy bombardment could replenish the water.

Overall, the reviews by Scalo et al (2007), Shields, Ballard, & Johnson (2016), and Kaltenegger (2017) are favorable towards these stars and their planets. Quoting Shields et al. directly, "While some characteristics of the M-dwarf stellar and planetary environment are still concerning, and more work remains to be done in the future to fully constrain their impact on habitability, some of these effects now indicate advantages for habitability, and serve as reasons to prioritize M-dwarf planets as targets in the search for the next planet where life exists."

**Synergies with JWST and Future Ground-Based Facilities**

JWST will certainly perform a reconnaissance of the most promising rocky exoplanets transiting mid-to-late M-dwarfs, primarily in transmission at near-infrared wavelengths (Morley et al. 2017). It is unlikely that *JWST+MIRI/LRS* (5 - 12 μm) will have the stability and precision necessary to obtain reliable mid-infrared spectra of rocky HZ planets (e.g., Greene et al. 2016), let alone detect biosignature gases such as $O_3$ and $CH_4$. Beyond 12 μm, MIRI will be limited to broadband photometry. From the ground, the next generation of extremely large telescopes will search for $O_2$ using high-resolution spectrographs (Snellen et al. 2013, Rodler et al. 2014). Overall, the landscape of *OST* will likely be one where we will have some knowledge about the targets of interest, such as whether or not a planet has an atmosphere or contains an indeterminate amount of $O_2$, but the picture will be incomplete. *OST* will determine the atmospheric compositions and thermal structures of dozens of potentially habitable planets, thus opening the door for comparative exoplanetology of rocky worlds.

**Atmospheric Characterization Across the Continuum of Cool Transiting Exoplanets**

The last decade of exoplanet science has shown us there is a broad continuum of planets across a factor of a thousand in mass from sub-Earths to super-Jupiters. *OST* would revolutionize our understanding of these worlds, especially those cooler than the planets that will be well-characterized in the near-infrared with *JWST*. For cooler planets, the abundances of $CH_4$ and $NH_3$ probe metallicity, non-equilibrium chemistry, and the strength of vertical mixing (Zahnle & Marley, 2014). The rich photochemistry expected for these atmospheres, via the production of HCN, $C_2H_2$, $C_2H_4$, and $C_2H_6$, can be probed by OST beyond 5 μm. *OST* can also advance our understanding of the physics and chemistry of clouds, which are currently a major source of uncertainty in models of these atmosphere.. A number of cloud species that are known or expected to impact the spectra of these planets over a wide temperature range have Mie scattering features in the mid-infrared (Wakeford & Sing, 2015; see Fraine et al. white paper).

Additionally, eclipse maps and phase curves will provide an unprecedented view of dynamics and circulation in cool atmospheres, as a bridge to the Solar System's gas giants.

**Direct Imaging of Jupiter-Sized Exoplanets**

With contrast ratios of $10^{-4}$ to $10^{-5}$, *OST* will be able to directly image and characterize the atmospheres of gas giant planets on wide orbits, critical for understanding the origin and evolution of exoplanetary systems like our own. We can assess the frequency and evolution history of wide-separation giant planets down to Saturn-mass. For additional information, see the white paper contribution by Meshkat, Nielsen et al. (2018).

**Technology Advancements**

Every future flagship mission needs technology advancement to make its exoplanet science case a reality. *OST* is no different. The *MISC* instrument is being designed in collaboration with JAXA and will use a densified pupil spectrometer (i.e., Matsuo et al. 2016) to generate multiple spectra on the detector plane that are stable against telescope pointing jitter and deformation of the primary mirror. Photon-noise-limited performance can be achieved by applying Independent Component Analysis (ICA) to these spectra. *MISC* will use state-of-the-art detectors to provide simultaneous wavelength coverage from 5 to 25 μm at R = 100 - 300. The success of the biosignatures science case relies on coadding spectra and a low instrument noise floor (~5 ppm precision) to build up sufficient signal-to-noise. High-stability detectors suitable for the mid-IR are already under investigation. We note that the cost to increase the test readiness level (TRL) of mid-IR detectors is relatively low compared to the overall cost of the mission.

**Conclusions**

*OST* is a mission concept that will provide new and fundamental insights into a wide range of cool exoplanetary atmospheres and *MISC*'s broad wavelength coverage will enable serendipitous discoveries that lead to new lines of scientific inquiry. *OST* will provide our first complete view of the atmospheric composition and temperature structure of Earth-size HZ planets and assess their ability to host life. Thus, for the first time in human history, we will have the means to answer one of our longest-standing questions: **"Are we alone?"**